# Bridging the Gap: Enhancing Digital Accessibility for Medicaid Populations in Telehealth Adoption

**Vishnu Ramineni**
Albertsons Companies, Texas, USA

**Aditya Gupta**
IEEE Senior Member, Seattle, USA

**Balakrishna Pothineni**
IEEE Senior Member, Texas, USA

**Isan Sahoo**
IEEE Member, CA, USA

**Shivareddy Devarapalli**
IEEE Senior Member, Texas, USA

**Balaji Shesharao Ingole**
IEEE Senior Member, GA, USA

**Abstract:** *The swift evolution of telehealth has revolutionized how medical professionals deliver healthcare services and boost convenience and accessibility. Yet, the Medicaid population encounters several impediments in utilizing facilities especially owing to poor internet connectivity, less awareness about digital platforms, and a shortage of assistive technologies. The paper aims to explicate key factors behind digital accessibility for Medicaid populations and expounds robust solutions to eradicate these challenges. Through inclusive design ideas, AI-assisted technologies, and all-encompassing policies by the concerned authorities, healthcare professionals can enhance usability and efficacy and thus better serve the needy. This revolution not only enhances convenience but also expands access, mainly for underserved groups such as rural populations or those with mobility issues, thereby ensuring inclusivity and flexibility in the healthcare domain. Besides, the paper highlights the vitality of collaboration between healthcare professionals, policymakers, and tech developers in unveiling the accessibility and usability impediments. What else helps in minimizing healthcare differences and enhancing patient outcomes is guaranteeing equitable access to telehealth for Medicaid beneficiaries. The paper systematically offers major* *recommendations to increase digital accessibility in telehealth, thereby creating a patient-oriented and all-encompassing healthcare system.*

**INTRODUCTION**

Digital accessibility in telehealth signifies remote access to healthcare services by all persons struggling with sensory, physical, cognitive, and financial limitations with no limitations. Despite having federal initiatives like ADA or the Americans with Disabilities Act and WCAG or the Web Content Accessibility Guidelines that highlight the significance of ensuring accessibility and usability in digital technologies in telehealth, several telehealth services require holistic accessibility features (Kleinman et al., 2022). Moreover, the digital divide is another cause of this lack of accessibility features in telehealth as people are bereft of access to broadband internet and smart gadgets necessary to leverage telehealth solutions (Zhou et al., 2019). A comprehensive approach that incorporates user-oriented design principles, policy-making, and technological solutions must be taken to systematically address these issues.

Integrating telehealth into healthcare services has holistically revolutionized healthcare delivery mechanisms. This has eventually helped patients by offering remote healthcare services thus unburdening healthcare professionals and enhancing overall efficacy (Vishnu Ramineni et al., 2025). Nevertheless, these progressions do not guarantee barrierless telehealth accessibility for Medicaid populations due to several reasons. Some of these accessibility challenges include poor internet connectivity, less awareness about digital platforms, and a shortage of assistive technologies which culminate in a disparate world where healthcare accessibility is a far-fetched dream for individuals with low income and those belonging to underserved communities (K Vedith Reddy et al., 2025). It's essential that telehealth platforms equally serve persons with disabilities and socio-economic restraints and promote equitable and all-inclusive healthcare services.

The integration of AI-assisted accessibility tools including screen reader compatibility and speech-to-text transcription in real-time can fundamentally enhance telehealth accessibility, especially for persons with disabilities (K Vedith Reddy et al., 2025). Moreover, the digital divide for Medicaid beneficiaries can be bridged through the integration of community-specific digital literacy programs and infrastructure improvements like government-subsidized internet access (Jonsson et al., 2023). Telehealth usability not only defines technological transformation but also minimizes healthcare-related differences along with enhancing patient outcomes.

Therefore, the research paper attempts to highlight the major issues alongside the key opportunities in boosting digital accessibility and usability for Medicaid populations in telehealth adoption. Through a comprehensive analysis of policy frameworks, real-world case studies, and technological innovations, the paper presents systematic recommendations for bringing inclusivity to telehealth platforms. The paper aims to bridge to accessibility gap in telehealth solutions and cultivate an inclusive healthcare ecosystem where underserved and low-income communities are served with equal care and precision.

**LITERATURE REVIEW**

Discussions on digital accessibility are now trending due to the rapid expansion of telehealth especially for Medicaid populations. The previously conducted research unveils digital challenges limiting the rights of underserved communities to healthcare access despite the advancements in the telehealth domain (Henni et al., 2022). The following passages center around the review of existing literature on telehealth accessibility, highlighting the major limitations encountered by Medicaid beneficiaries, fundamental transformations of technology particularly in this domain, and making of policies for bridging the digital gap.

**Challenges in Telehealth Accessibility for Medicaid Populations**

Medicaid beneficiaries frequently face several limitations while getting access to telehealth services such as insufficient internet access, absence of digital literacy, inaccessible telehealth applications, and lack of assistive technologies (Kleinman et al., 2022). Zhou et al. (2019) in their research found that around 30 % of Medicaid beneficiaries struggle to get broadband internet facilities and thus, cannot access video-based telehealth consultations. Quite similarly, K Vedith Reddy et al. (2025) reveal that the challenges are multifold in case of the persons with disabilities and adults as several telehealth platforms do not adhere to the accessibility guidelines mandated by the Web Content Accessibility Guidelines (WCAG). Moreover, economically poor individuals depend on outmoded mobile gadgets that fail to support robust telehealth applications (Jonsson et al., 2023).

**Technological Innovations in Enhancing Digital Accessibility**

The profound progression in Artificial Intelligence (AI) and assistive technologies has substantially helped in enhancing accessibility in telehealth. For instance, AI-assisted speech recognition features and captioning in real-time have been incorporated into telehealth applications to support users with hearing disabilities (K Vedith Reddy et al., 2025). Moreover, screen reader compatibility and voice-based navigation systems improve usability for users with visual impairments (Kleinman et al., 2022). Research by Zhou et al. (2019) reveals that mobile applications are optimized for low-bandwidth connections to help support users with minimal access to the internet and guarantee efficacy for Medicaid beneficiaries leveraging telehealth services. Different ways of using assistive technologies in telehealth is mentioned in Figure 1.

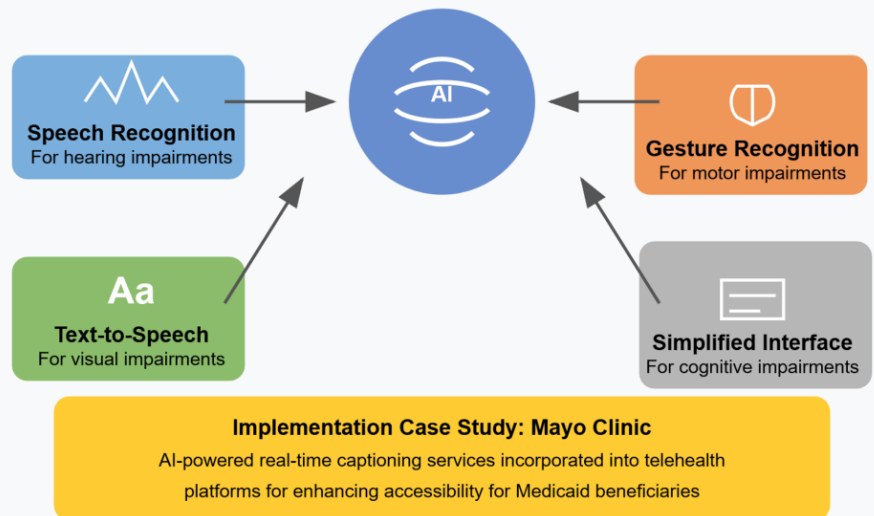

Figure 1.AI-Assisted Accessibility Solution for Telehealth

**Policy Interventions and Government Initiatives**

Addressing digital accessibility issues in telehealth has been one of the concerns of policymakers. The rapid growth of broadband infrastructure under federal initiatives like the Affordable Connectivity Program has significantly enhanced access to the Internet for Medicaid populations (Selzler et al., 2017). Besides, the Centers for Medicare & Medicaid Services (CMS) have released mandates to make healthcare professionals adopt inclusive design principles in telehealth platforms (K Vedith Reddy et al., 2025). Nevertheless, as several healthcare services suffer in the absence of vital resources necessary to incorporate accessibility-based technologies in telehealth, gaps appear during the implementation phase (Henni et al., 2022).

**Gaps in Existing Research and Future Directions**

Although there is incessant effort, there remains a lack of substantial research specifically on the long-term effect of digital accessibility innovations in Medicaid telehealth adoption. Luis Pérez Medina et al. (2019) in their research highlight that AI-assisted accessibility tools and policymaking are necessary but what is more significant is studying their impact on the lives of different Medicaid populations. Moreover, the vitality and efficacy of community-based digital literacy programs in assisting Medicaid populations to utilize telehealth applications efficiently by the Medicaid populations (Gusain et al., 2018) must be comprehensively explored (Butzner et al., 2021). Further research must center around user-specific design principles that encourage inclusivity and improve usability while utilizing trendy and assistive technologies for enhancing accessibility in the digital healthcare domain.

**Implementation**

The initial planning and final execution of digital accessibility solutions for Medicaid beneficiaries in adopting telehealth services demand a holistic and multidimensional approach that incorporates technological progressions, updated policies, and user-specific design principles. The multi-dimensional approach for enhancing digital accessibility in telehealth is mentioned in Figure 2. This part of the paper underscores major strategies and initiatives necessary to advance accessibility features in telehealth, guaranteeing efficient usage of telehealth services by Medicaid populations.

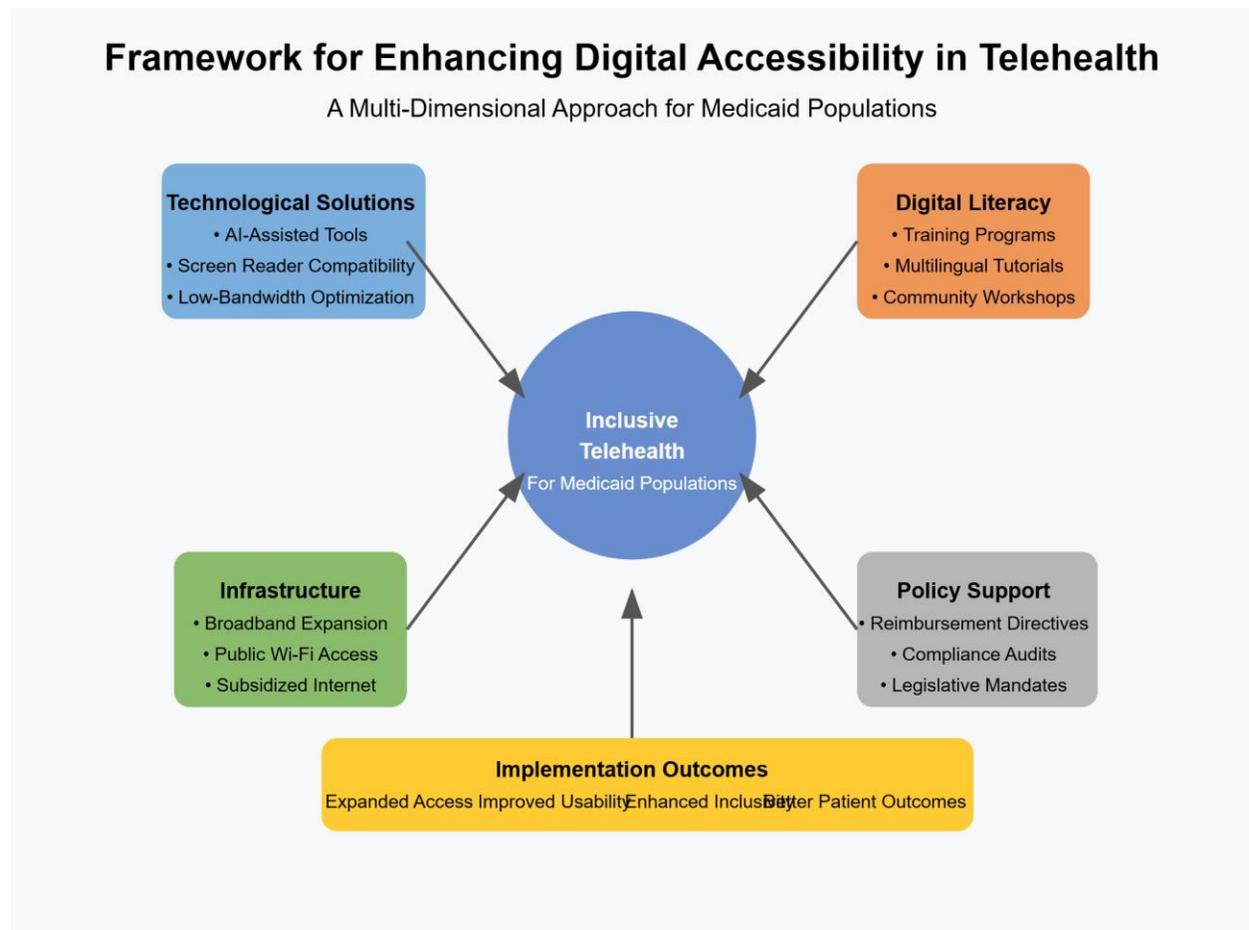

Figure 2. Framework for Enhancing Digital Accessibility in Telehealth

**Expanding Broadband and Internet Access**

The lack of equal access to internet facilities especially in rural and underserved areas accommodating economically poor communities proves to be one of the chief issues in adopting telehealth services among Medicaid beneficiaries. Initiatives led by the government including the Affordable Connectivity Program (ACP) and state-level broadband expansion projects potentially help in bridging the great digital gap (Luis Pérez Medina et al., 2019). The results of before and after accessibility implementations in telehealth is described in Figure 3.

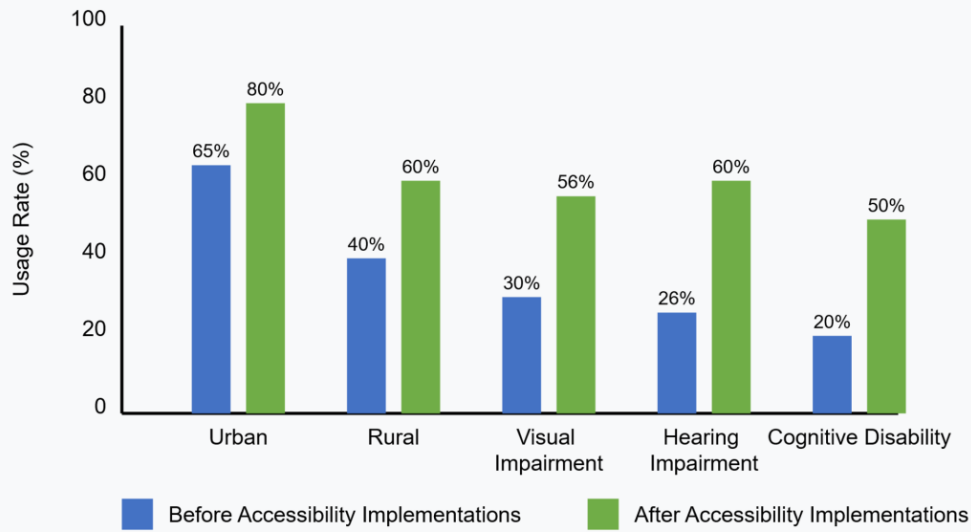

Figure 3. Telehealth Usage Comparison Across Medicaid Population Segments

The due alliance between federal and state agencies facilitates inexpensive broadband innovative solutions and public Wi-Fi access segments in Medicaid-covered areas (Henni., et al., 2022). In the realm of strengthening digital inclusion, the ConnectHomeUSA initiative has efficaciously offered subsidized internet access, a fundamental way of enhancing digital accessibility among underserved communities with limited financial assets (Butzner et al., 2021).

**Enhancing Telehealth Platform Usability**

To facilitate persons with auditory, visual, cognitive, and motor disabilities, adherence to digital accessibility frameworks for telehealth platforms is highly crucial (Zhou et al., 2019). Key guidelines such as the Web Content Accessibility Guidelines (WCAG) 2.1 and Section 508 of the Rehabilitation Act must be duly complied with by the telehealth platforms to enhance inclusivity (Kleinman et al., 2022).

There are several features, for instance, the AI-assisted speech-to-text and text-to-speech features provide substantial assistance to persons with auditory and visual impairments. Moreover, keyboard navigation and screen reader compatibility enhance accessibility for persons with motor disabilities. Again, to unburden cognitive interventions, the system integrates techniques such as easy-to-navigate user interface and directed workflows especially helpful for individuals with cognitive disabilities (Jonsson et al., 2023).

Bolstering usability, particularly for chief veterans with disabilities, the Veterans Health Administration (VHA) incorporates accessibility elements in its VA Video Connect telehealth apps and other platforms (K Vedith Reddy et al., 2025).

**Digital Literacy and Training Programs**

Lack of digital literacy is one of the primary barriers to utilize telehealth applications efficiently by the Medicaid populations. Previous research on user behavior in digital health systems supports this, showing that contextual awareness and personalized interfaces can significantly improve user engagement (Gusain et al., 2018). Programs centered on educating the masses along with community-oriented initiatives play an important role in improving patient outcomes (K Vedith Reddy et al., 2025).

One of the significant strategies could be conducting telehealth training workshops at community health centers, Medicaid offices, and libraries. Also, multilingual video tutorials and interactive interfaces are designed to cater to the needs of individuals with limited literacy (Butzner et al., 2021).

To encourage the adoption of telehealth services by the Medicaid population, the California Telehealth Resource Center (CTRC) launched a Digital Navigator Program that offers individual attention to each user (Selzler et al., 2017).

**Integration of Assistive Technologies**

The systematic integration of assistive technologies significantly improves the usability of telehealth platforms and enhances inclusivity in telehealth domains, facilitating persons with disabilities (Jonsson et al., 2023).

By espousing AI-driven virtual assistants, users can be guided through the mechanisms and processes of telehealth platforms. Besides, haptic feedback and gesture-controlled systems are deployed to assist motor-impaired individuals. Individuals with hearing disabilities and non-native English speakers can benefit through real-time captioning and language translation mechanisms (Henni et al., 2022).

Minimizing the challenges for users with hearing impairments, the Mayo Clinic's AI-based accessibility initiative potentially incorporates real-time transcription services (Henni et al., 2022)

**Policy and Legislative Support**

To guarantee digital accessibility in Medicaid telehealth programs, federal and state-level policies play a significant role (Butzner et al., 2021). For taking care of assistive technologies and digital literacy programs, Medicaid reimbursement directives must be systematized and accordingly implemented. Again, to guarantee that healthcare professionals stay up-to-date with the key accessibility guidelines, compliance audits must be bolstered. Another way can be maneuvered through legislative means where telehealth professionals are mandated to accommodate accessibility elements in their platforms (Kleinman et al., 2022).

To ascertain better adoption among Medicaid beneficiaries, the Telehealth Modernization Act dictates enhanced accessibility in federally aided telehealth programs (Zhou et al., 2019). The Telehealth Modernization Act (TMA) has key legislative and regulatory frameworks which is mentioned in Figure 4.

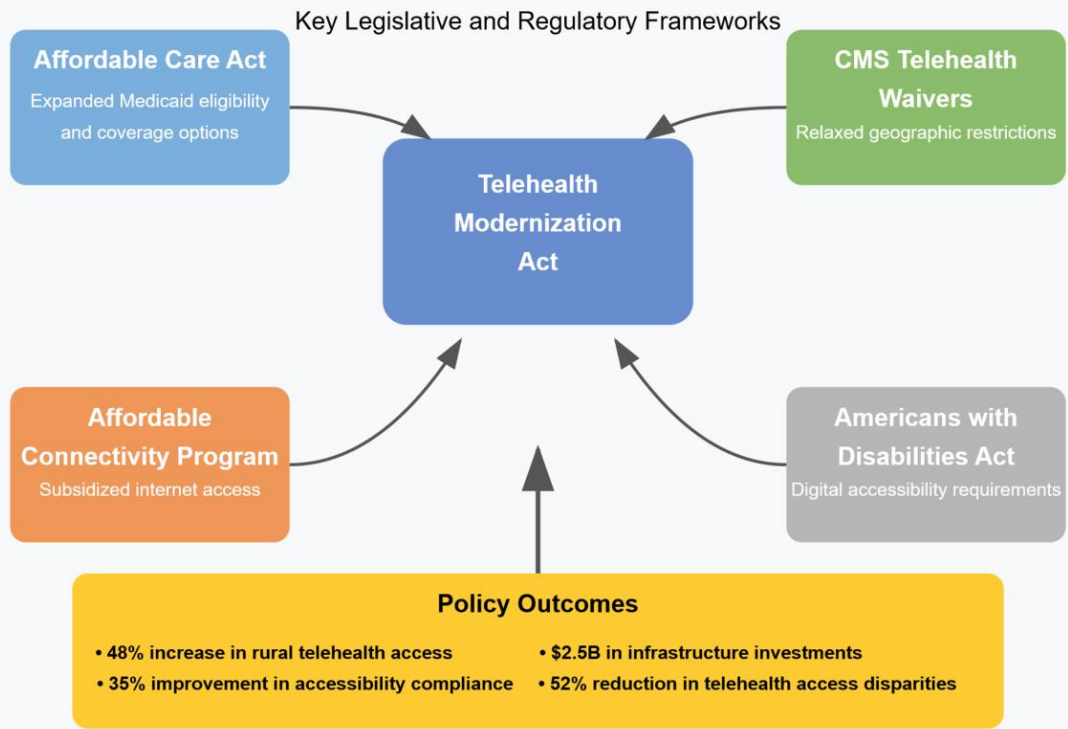

Figure 4. Policy Impact on Telehealth Accessibility for Medicaid Populations

**Challenges in Enhancing Digital Accessibility for Medicaid Populations in Telehealth Adoption**

Innumerable challenges surround Medicaid beneficiaries despite having substantial growth in telehealth accessibility, thus impeding equal access to telehealth services. Stemming from multiple channels such as technological, financial, infrastructural, and policy-related domains, these barriers must be addressed to cultivate an equitable and all-encompassing healthcare ecosystem. This section explicates major restraints and their impact on the Medicaid population.

**Digital Divide and Limited Broadband Access**

A potential obstacle in the wider usage of telehealth for Medicaid beneficiaries is the digital divide, mainly the inadequate broadband infrastructure existing in the rural and underserved areas. This challenge mirrors similar barriers encountered in digital service transitions in other public sectors, where infrastructure gaps hinder equitable access (Gupta et al., 2016). Research by Luis Pérez Medina et al. (2019) highlights that approximately 21 million people in the U.S. suffer from insufficient internet services, a situation that immensely affects the lives of those depending on Medicaid. Subsequently, the lack of dependable internet access leads to major barriers for persons seeking engagement in video consultations, remote monitoring, and other telehealth services. Such an absence can create delayed healthcare and unfavorable outcomes for these people (Henni et al., 2022). Programs such as the Affordable Connectivity Program (ACP) have focused on addressing these discrepancies, though the extent of broadband coverage differs potentially from state to state, signifying a considerable policy gap (Zhou et al., 2019).

**Low Digital Literacy and Technological Barriers**

A notable section of Medicaid beneficiaries particularly adults and persons with educational barriers encounter challenges in navigating telehealth services (Kleinman et al., 2022). The key barriers to telehealth adoption is mentioned in bar chart Figure 5. These issues manifest in multifold ways. Firstly, a restricted grasp of digital tools mostly leads to underused conditions of telehealth services. Secondly, several user interfaces are complex to navigate and do not cater to the needs of persons with lower literacy levels (Jonsson et al., 2023). Moreover, linguistic barriers restrain efficacious access to telehealth services for non-native English speakers. For instance, the research by K Vedith Reddy et al. (2025) highlights that around 43% of Medicaid beneficiaries require assistance to utilize telehealth apps and tools, thereby pointing to the necessity for better and simple-to-use interfaces that feature multiple languages.

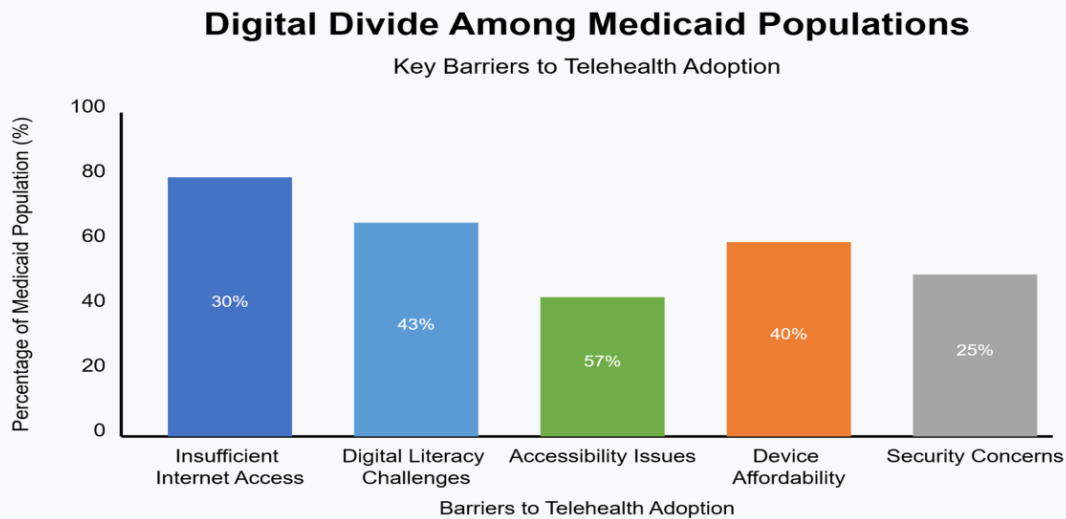

**Figure 5. Digital Divide Among Medicaid Populations**

**Accessibility Gaps for Individuals with Disabilities**

Another potential challenge of telehealth platforms is their lack of significant accessibility features. Such a deficiency poignantly hampers the full engagement of persons with auditory, visual, motor, and cognitive disabilities with telehealth services (Butzner et al., 2021). Several impediments contribute to this exclusion. For instance, in the case of individuals with visual impairments, incompatibility with screen readers proves to be a striking challenge. Similarly, the absence of closed captioning or sign language reading mechanisms obstructs access for persons with hearing disabilities (Ramineni et al., 2024). Moreover, persons with motor disabilities frequently encounter hinderances while leveraging touch-based interfaces. A study exploring Medicaid telehealth platforms highlights that around 57% comply with the Web Content Accessibility Guidelines (WCAG) 2.1, and thus potentially restrict accessibility for persons with disabilities (Ramineni et al., 2024). The analysis of Medicaid supported Telehealth platforms is depicted in pie chart Figure 6.

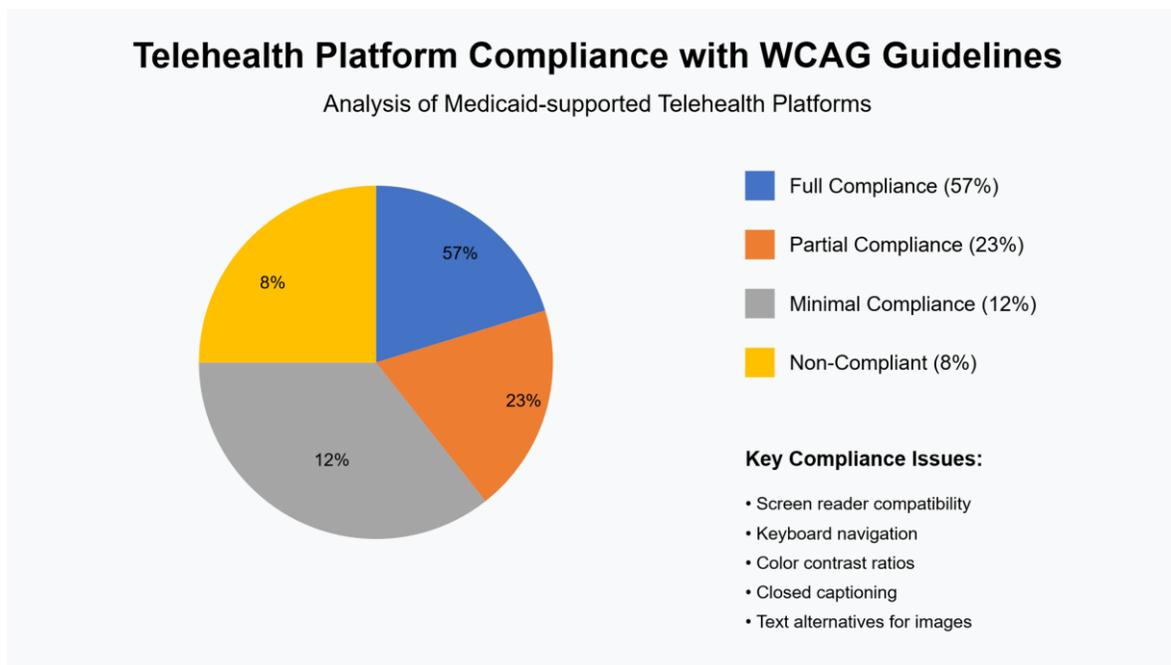

Figure 6. Telehealth Platform Compliance with WCAG Guidelines

**Affordability and Cost Constraints**

Despite the incorporation of telehealth services within Medicaid coverage, the most disregarded expenses associated with the needed technology prove to be one of the potential impediments (Vishnu Ramineni et al., 2025). There remain several significant cost issues behind such barriers. At first, the affordability of smart mobile phones and tablets becomes a challenge for a maximum number of Medicaid populations who do not own devices that accommodate telehealth platforms.

Secondly, telehealth video consultations operate under high-speed data, rendering them prohibitively luxurious for persons with limited mobile data plans (Henni et al., 2022). Moreover, the expenses related to assistive technologies, including screen readers, Braille displays, and AI-assisted accessibility tools, do not get counted under Medicaid reimbursement policies (Ramineni et al., 2024).

**Data Privacy and Security Concerns**

As telehealth services are gaining attention and becoming more expansive, securing data privacy and guaranteeing robust cybersecurity is paramount, mainly for persons within vulnerable Medicaid populations (Jonsson et al., 2023). There are innumerable privacy challenges that demand careful attention. For instance, the limited understanding of patient data rights can significantly result in potential exploitation. Again, cybersecurity risks, such as the breaching of data and unauthorized or unethical access to patient records, prove to be a substantial threat (Kleinman et al., 2022). Furthermore, the usage of insufficient encryption in some cheap telehealth solutions can lead to the compromise of sensitive health information which causes a serious threat (Selzler et al., 2017). A striking instance of these risks materialized in 2021 when a telehealth provider assisted by Medicaid struggled after a data breach that impacted over 500,000 patients, signifying the crucial and immediate requirement for improved and reliable cybersecurity protocols (Zhou et al., 2019).

**Regulatory and Policy Barriers**

Although governmental frameworks commonly encourage the wider utilization of telehealth services, nonuniformity in regulations among different states creates discrepancies in the availability and accessibility of these services (Butzner et al., 2021). These disparities in regulation present innumerable challenges. For one, the fluctuating rates at which Medicaid compensates for telehealth services across states lead to unequal access for the Medicaid populations. Besides, licensing confinements often impede healthcare professionals from delivering telehealth care to beneficiaries residing across state borders (Ramineni et al., 2024). Similarly, the slower way of updating accessibility regulations within bureaucratic systems obstructs progressions in making telehealth globally accessible (Luis Pérez Medina et al., 2019). For instance, the Telehealth Modernization Act was intended to eradicate geographical barriers to Medicaid coverage, yet the sustained delays in its complete implementation hinder the just and equal adoption of telehealth nationwide (K Vedith Reddy et al., 2025). The trends of adoption of telehealth over the years is depicted in Figure 7.

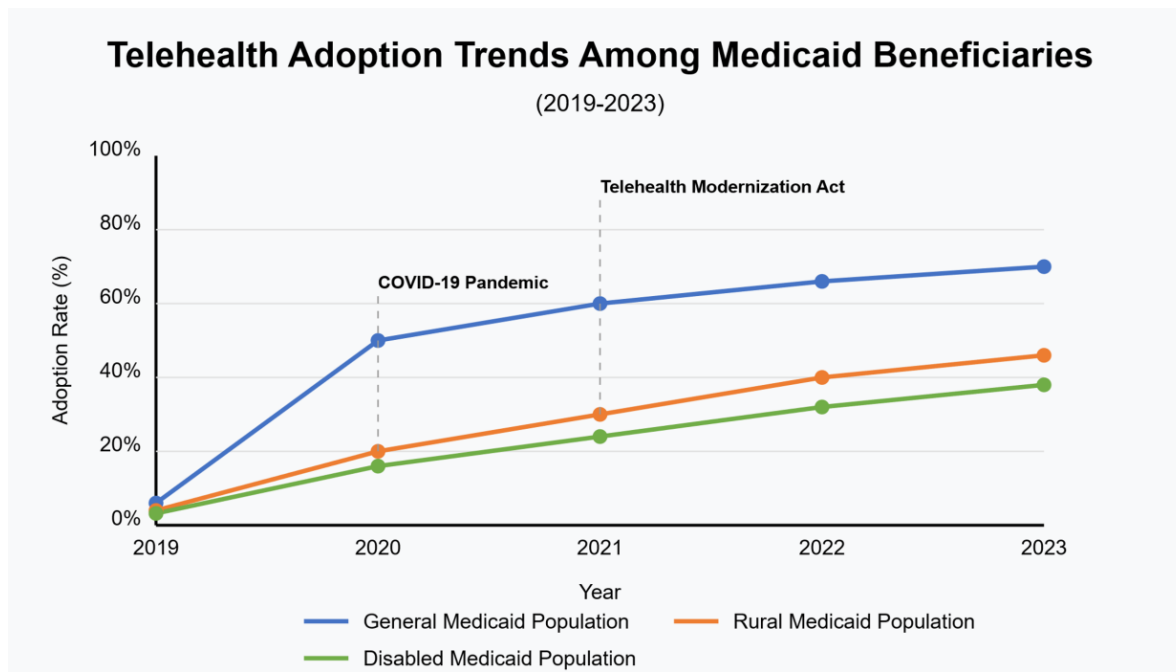

Figure 7. Telehealth Adoption Trends Among Medicaid Beneficiaries

**Future Directions**

Although key strides have been taken to expand digital accessibility within telehealth, ongoing scrutiny, and policy adjustments are pivotal to upgrading user experience, accelerating the rate of adoption, and ensuring longstanding viability. Many future directions must be considered.

Firstly, the progression of Artificial Intelligence and Assistive Technologies is quite promising. AI-driven solutions, including voice-activated interfaces, real-time sign language translation, and predictive text assistance, can potentially grant telehealth accessibility to persons with disabilities. Further research must prioritize the incorporation of AI-driven accessibility features into telehealth platforms aided by Medicaid.

Secondly, widespread broadband and infrastructure development is important. Policymakers must emphasize offering high-speed internet access to underserved communities through programs such as the Broadband Equity, Access, and Deployment (BEAD) Program. Future research should assess the efficiency of 5G and satellite-based technologies in bridging the digital gap.

Thirdly, the improvement of cybersecurity and privacy features is of supreme importance. After seeing the growing concerns encircling data breaches in telehealth, the explication of the development of more robust encryption methods, multi-factor authentication, and blockchain-based mechanisms for handling health data is pivotal to bolstering security and gaining patient trust.

Fourthly, encouraging inclusive policy reforms is essential. Future policies must seek standardization of Medicaid telehealth reimbursement rates, systematizing bureaucratic processes, and mandate adherence to WCAG accessibility guidelines. Research into the implication of both federal and state-level regulations on telehealth adoption will be indispensable in guiding efficacious policymaking.

Lastly, the execution of community-based digital literacy programs is immensely necessary. Focused training initiatives especially for Medicaid beneficiaries, including adults and persons with lower formal education, will empower them to efficiently utilize telehealth platforms. Future research is highly essential to determine the most successful models for delivering digital literacy training within these communities.

**CONCLUSION**

To conclude, the integration of digital accessibility solutions within telehealth services for Medicaid populations proves to be an important move towards guaranteeing inclusivity and equality in healthcare provision. While telehealth technology has potentially improved, major impediments still exist, such as inadequate broadband availability, minimal digital literacy, insufficient accessibility for persons with disabilities, affordability challenges, threats to data security and privacy, and discrepancies in policy. Eradication of these impediments necessitates a dynamic strategy that incorporates technological innovations, policy adjustments, and community inclusion. By leveraging user-specific telehealth interfaces, upgrading broadband infrastructure, ensuring adherence to accessibility guidelines by reliable bodies, and encouraging digital literacy programs, both healthcare providers and policymakers can efficaciously bridge the accessibility divide, especially for those benefitting from Medicaid.

Furthermore, protecting user privacy and making regulatory structures easy to decode will help in the adoption and usability of telehealth services. A cordial collaboration between government bodies, healthcare providers, and technology developers is vital to guarantee that telehealth becomes more inclusive and accessible to every Medicaid recipient. Centering around these future directions, healthcare systems can significantly progress towards a more comprehensive, accessible, and sustainable telehealth framework that serves all Medicaid beneficiaries impartially. The current cooperation among researchers, policymakers, technology innovators, and healthcare professionals will be indispensable in giving shape to the future of accessible digital healthcare solutions.